\documentclass{ws-ijmpa}
\usepackage[super,compress]{cite}
\usepackage{graphicx}
\usepackage{pifont}

\usepackage{hyperref}
\usepackage[caption=false]{subfig}

\begin{document}
\markboth{T.~Bla\v zek, P.~Mat\'ak}
{Left-left squark mixing in $K^+\rightarrow\pi^+\nu\bar{\nu}$ and minimal SUSY}

%
\catchline{}{}{}{}{}
%

\title{LEFT-LEFT SQUARK MIXING, $K^+\rightarrow \pi^+\nu\bar{\nu}$ AND MINIMAL SUPERSYMMETRY WITH LARGE $\tan\beta$}

\author{TOM\'A\v S BLA\v ZEK}

\address{Department of Theoretical Physics, Comenius University, \\
Mlynsk\' a Dolina, Bratislava, 84248,\\
blazek@fmph.uniba.sk}

\author{PETER MAT\'AK}

\address{Department of Theoretical Physics, Comenius University, \\
Mlynsk\' a Dolina, Bratislava, 84248,\\
peter.matak@fmph.uniba.sk}

\maketitle

\begin{history}
\received{Day Month Year}
\revised{Day Month Year}
\end{history}

\begin{abstract}
We analyze the role of the left-left squark mixing in the rare $K^+ \rightarrow \pi^+\nu\bar{\nu}$ decay within the minimal supersymmetry with a large $\tan\beta$. A Higgs boson mass of $125$ GeV has been taken into account leading to correlation between stop masses and trilinear soft supersymmetry breaking coupling $A_{\tilde{t}}$. We find that measurable effects, similar to that of the well known $LR$ squark mixing terms, are possible for large $A_{\tilde{t}}$ combined with the off-diagonal $LL$-insertions. Precise measurements of the decay rate are expected from the ongoing NA62 experiment at CERN. We emphasize that the effect we present can put certain limits on the left-left flavor changing structure of the squark mass matrix.

\keywords{kaon decays; supersymmetry; non-minimal flavor violation.}
\end{abstract}

\ccode{PACS numbers: 11.30.Pb; 13.20.Eb.}


\section{Introduction}\label{intro}

Flavor Changing Neutral Currents (FCNC) form special class of decays and meson mixings that are very suitable for indirect testing of the new physics models. In the Standard Model (SM) they arise at loop level only, which leads to the suppression of their amplitudes. On the other hand, beyond Standard Model scenarios usually contain new particles or an extended Higgs sector, both able to affect the processes in a significant way. Flavor changing processes are often sensitive to the predictions of GUTs. In the supersymmetric versions, the ratios of the gaugino masses are fixed, leading to heavy gluinos and suppression of their flavor changing diagrams. The $\mathrm{SO}(10)$-like unification posses a large value of $\tan\beta\approx 50$, which enhances the higgsino couplings having an impact on the FCNC amplitude. 

However, most of these processes suffer from large hadronic uncertainties due to the nature of the initial states. Among the FCNC, the rare kaon decays, $K^+\rightarrow \pi^+\nu\bar{\nu}$ and $K_L\rightarrow \pi^0\nu\bar{\nu}$, can be calculated with the exceptional precision. Their hadronic matrix elements can be extracted from the semileptonic kaon decay, $K^+\rightarrow\pi^0e^+\nu_e$ \cite{Gaillard:1974hs}. To reduce the systematic errors, isospin breaking corrections and radiative QED corrections \cite{Mescia:2007kn}, long distance effects and higher dimensional effective operators \cite{Isidori:2005xm} were considered. With the next-to-next-to-leading order QCD corrections to the charm quark contribution computed in Ref. \citen{Buras:2005gr}, the main source of the systematic error remain the uncertainty in the charm quark mass $m_c$ and the CKM matrix elements \cite{Buras:2005gr}. As anticipated in Ref. \citen{Buras:2004uu}, future improvement of the $m_c$ measurement can lead to the $BR(K^+\rightarrow\pi^+\nu\bar{\nu})$ prediction with the theoretical uncertainty as small as $\pm5\%$.

It is not a surprise that such theoretically clean decay attracts the attention of experimental physicists. So far, seven $K^+\rightarrow\pi^+\nu\bar{\nu}$ events have been collected by E-787, E-949 experiments of Brookhaven National Laboratory \cite{Adler:2000by,Anisimovsky:2004hr,Artamonov:2008qb}, leading to
\begin{equation}
BR(K^+\rightarrow\pi^+\nu\bar{\nu})_{exp} = (1.73^{+1.15}_{-1.05})\times 10^{-10}.
\end{equation}
In order to obtain precise measurement of the decay branching ratio, the NA62 experiment was designed at CERN, starting to collect the data in the autumn this year. The aim of this experiment is the collection of about 100 events of $K^+\rightarrow\pi^+\nu\bar{\nu}$ for the Standard Model decay rate \cite{Ruggiero2011216}, allowing to probe the beyond Standard Model scenarios and put limits on their flavor breaking parameters.

\section{\texorpdfstring{$K^+\rightarrow \pi^+\nu\bar{\nu}$}{Charged kaon decay}: Standard Model and beyond}

\subsection{From Effective Lagrangian to Branching Ratio}

In the standard model, the $K^+\rightarrow \pi^+\nu\bar{\nu}$ decay amplitude is dominated by $D=6$ effective operator
\begin{equation}
\mathcal{O}_{L}=(\bar{s}\gamma^\mu P_{L} d)(\bar{\nu}_{l}\gamma_\mu P_L\nu_{l}),
\label{op}
\end{equation}
with the corresponding Wilson coefficient having contribution from the ten penguin and four box diagrams. In the `tHooft-Feynman gauge the dominant effect comes from the penguins and box with top and two $W^\pm$. Supersymmetric new physics scenarios extend the possible sources of the flavor violation through the squark mass matrices. Thanks to the enlarged Higgs sector, a non-zero effect of the right-handed quark current occurs. Therefore, the MSSM effective Lagrangian can be written as \cite{Colangelo:1998pm,Isidori:2006jh,Buras:1998raa}
\begin{equation}
\mathcal{L}_{\mathrm{eff}}=\mathcal{L}_{c,\mathrm{eff}}+\frac{4 G_F}{\sqrt{2}}\frac{\alpha}{2\pi \sin^2\theta_W} \lambda_t
\sum\limits_{l=e,\mu,\tau}\big[X_L \mathcal{O}_L+X_R \mathcal{O}_R\big].
\label{lag}
\end{equation}
In the above formula, $\mathcal{L}_{c,\mathrm{eff}}$ includes physics below the electroweak scale and is fully dominated by the standard model charm quark loops \cite{Buras:2005gr}. In our notation $\lambda_q =V_{qd} V^*_{qs}$, $\lambda =|V_{us}|$. Although the corresponding loop functions depends on quark mass approximately as $\propto m^2_q/M^2_W$, the charm loops cannot be neglected due to the significantly larger CKM factor \cite{Isidori:2005xm}.

As we already mentioned, the great advantage of the $K^+\rightarrow\pi^+\nu\bar{\nu}$ decay is the fact that the non-perturbative matrix element of the quark current can be extracted from the measurement of the tree level semileptonic $K^+ \rightarrow \pi^0 e^+ \nu_e$ decay \cite{Gaillard:1974hs}. Using the strong isospin symmetry we obtain approximate relation
\begin{equation}
\langle \pi^+\nu\bar{\nu}\vert\bar{s}\gamma_\mu P_{L,R} d\vert K^+\rangle \doteq \sqrt{2}\langle\pi^0 e^+\nu_e\vert \bar{s}\gamma_\mu P_L u\vert K^+\rangle,
\label{HadrElem}
\end{equation}
where, as far as the electron is treated massless, the effect of the leptonic current is the same for both decays. After the NLO isospin breaking corrections are included, the hadronic matrix elements enter the $K^+\rightarrow\pi^+\nu\bar{\nu}$ branching ratio through the parameter $\kappa_+$. Then \cite{Colangelo:1998pm,Buras:1998raa}
\begin{equation}
\mathrm{BR}(K^+\rightarrow\pi^+\nu\bar{\nu})= \kappa_+ \bigg[
\bigg(\frac{\mathrm{Im}\lambda_t}{\lambda^5}X\bigg)^2 +\bigg(\frac{\mathrm{Re}\lambda_c}{\lambda} (P_c+\delta P_{c,u})+\frac{\mathrm{Re}\lambda_t}{\lambda^5}X\bigg)^2\bigg]
\end{equation}
and we quote the value of Ref.~\citen{Mescia:2007kn},
\begin{equation}
\kappa_+ = (5.173 \pm 0.025)\times 10^{-11} \times\big(\tfrac{\lambda}{0.2255}\big)^8.
\end{equation}

The charm contribution is not sensitive to the physics at the high energy scale and we use the NNLO result \cite{Brod:2008ss}
\begin{equation}
P_c = (0.372\pm 0.015) \times \big(\tfrac{0.2255}{\lambda}\big)^4,
\end{equation}
while long distance effects have been included in $\delta P_{c,u} = 0.04\pm 0.02$ \cite{Isidori:2005xm}.

In our notation $X=X_L+X_R$ stands for the top quark and short distance physics contributions. In the standard model $X^{SM}_R=0$ and $X^{SM}_L\simeq X_0(x_t)$, where $x_t=m^2_t/M^2_W$. The loop function $X_0(x_t)$ represents the sum of the Standard Model top quark diagrams and is equal to \cite{Inami:1980fz}
\begin{equation}
X_0(x_t)=\frac{x_t[x^2_t+x_t-2+3(x_t-2)\ln x_t]}{8(1-x_t)^2}.
\end{equation}
Inclusion of NLO QCD corrections \cite{Buras:2004uu} and two-loop electroweak \cite{Brod:2010hi} corrections lead to $X(x_t)=1.469\pm 0.02$ and resulting branching fraction \cite{Brod:2010hi}
\begin{equation}
BR(K^+\rightarrow\pi^+\nu\bar{\nu})_{SM} = (7.81^{+0.80}_{-0.71} \pm 0.29)\times 10^{-11}
\end{equation}
with the first error dominated by the uncertainty in the CKM matrix elements and the second coming from $P_c$ and $\kappa_+$ \cite{Brod:2010hi}.

\subsection{Supersymmetry and flavor Violation - definition of model}

Extending the standard model Lagrangian by supersymmetric and soft supersymmetry breaking terms naturally brings new sources of the flavor violation. It is no surprise that the resulting FCNC amplitudes depends on the sparticle masses as well as the pattern of their mixing. 

In our notation the bino, wino and gluino masses are $M_1$, $M_2$ and $M_3$, respectively, and fulfill GUT boundary condition
\begin{equation}
M_1:M_2:M_3 = \frac{5}{3}g^2_1:g^2_2:g^2_3.
\end{equation}
In order to obtain the chargino mass eigenstates, the wino-higgsino mass matrix has to be diagonalized as follows
\begin{eqnarray}
U^*\left(\begin{array}{cc}
M_2 & \sqrt{2}M_W s_\beta\\
\sqrt{2}M_W c_\beta & \mu
\end{array}\right)V^{-1}=\mathrm{diag}(M_{\tilde{\chi}_1},M_{\tilde{\chi}_2}).
\label{GUT}
\end{eqnarray}
Unitary matrices $U,V$ are of importance whereas they enter the $\tilde{\chi}_A-\tilde{u}_\alpha-d_i$ vertices\footnote{In the following, $U, V$ matrix elements are labeled by capital $A,B,\ldots$.}.

In the super-CKM\footnote{The basis in which squark fields are rotated by the same unitary matrices $V_{qL,R}$ that diagonalize Yukawa couplings and obey $V_{uL}V_{dL}^\dagger = V_{CKM}$.} basis, we assume the squark mass matrix of the form
\begin{eqnarray}
M^2_{\tilde{q}}=\left(
\begin{array}{ccc}
M^2_{\tilde{q},LL} && M^2_{\tilde{q},LR}\\
M^{2\dagger}_{\tilde{q},LR} && M^2_{\tilde{q},RR}
\end{array} \right)
\label{M2sq}
\end{eqnarray}
with\footnote{Notation: $T^3_q, Q_q$ - third isospin component and charge of the quark, $\mathbf{m}_q$ - diagonal quark mass matrices.}
\begin{eqnarray}
M^2_{\tilde{q},LL}  =  V_{qL}\mathbf{m}^2_{\tilde{Q}}V^{\dagger}_{qL}+\mathbf{m}^2_q+\big(T^3_q-Q_q s^2_W\big)M^2_Z\cos 2\beta \mathbf{1},
\label{MFVblocks1}\\
M^2_{\tilde{q},RR}  =  V_{qR}\mathbf{m}^2_{\tilde{q}}V^{\dagger}_{qR}+\mathbf{m}^2_q+Q_q s^2_W M^2_Z\cos 2\beta \mathbf{1},\\
M^2_{\tilde{q},LR}  =  (\mathbf{A}_{\tilde{q}}-\mu^*\cot\beta) \mathbf{m}_q.
\label{MFVblocks4}
\end{eqnarray}
Then, squark masses are obtained after the diagonalization of $M^2_{\tilde{q}}$ by means of unitary the transformation $R^{\tilde{q}}$,
\begin{equation}
R^{\tilde{q}} M^2_{\tilde{q}} R^{\tilde{q}\dagger}=\mathrm{diag}(M^2_{\tilde{q}_1}.\ldots,M^2_{\tilde{q}_6}).
\end{equation}
There are two types of contributions entering Eq.~(\ref{M2sq}) - the $SU(2)_L$ breaking terms proportional $v_{u,d}$ as well as explicit mass terms originating from the soft SUSY breaking part of the MSSM Lagrangian. While the first type is flavor diagonal or proportional to Yukawa couplings, the latter remains in general unrelated to the SM quantities and belongs to the 'yet to be measured in FCNC' list. However, mixing pattern of squarks, encoded in the soft SUSY mass terms $\mathbf{m}^2_{\tilde{Q}}$,$\mathbf{m}^2_{\tilde{q}}$  and $\mathbf{A}_{\tilde{q}}$ above, is crucial for all the flavor violating processes and can be constrained by several assumptions. 

Allowing for a general flavor structure, the soft SUSY breaking terms lead to new effects in FCNC amplitudes. Possible off-diagonal elements in the $XY=RR$ and $XY=LR$ blocks in Eq.~(\ref{M2sq}) are parametrized as dimensionless quantities
\begin{equation}
\delta^{ij}_{\tilde{q}XY}=\frac{(M^2_{\tilde{q},XY})^{ij}}{\sqrt{(M^2_{\tilde{q},XX})^{ii}(M^2_{\tilde{q},YY})^{jj}}}, i\neq j.
\label{deltaNMFV}
\end{equation}
In the $LL$ part of Eq.~(\ref{M2sq}) the situation is a bit more complicated. Since the left-handed squarks are members of the same electroweak doublet, in general, their mass matrices cannot be diagonalized simultaneously. There are two different ways, in which the flavor violation in the left-left squark sector can be introduced. If one assumes universal scalar masses at the GUT scale, the effect of the large top and bottom Yukawa coupling in the RGE will cause third family masses be small\-er than the first two. Therefore, as a starting point, we assume
\begin{equation}
\mathbf{m}^2_{\tilde{Q}}=\mathrm{diag}(m^2_{\tilde{Q}_1},m^2_{\tilde{Q}_1},m^2_{\tilde{Q}_3}), m^2_{\tilde{Q}_1}>m^2_{\tilde{Q}_3}
\end{equation}
and similarly for $\mathbf{m}^2_{\tilde{u},\tilde{d}}$. For the left-handed squarks usually the popular choice of the Buras-Romanino-Silvestrini basis\footnote{It becomes straightforward that, unless $m^2_{\tilde{Q},\tilde{u},\tilde{d}}\propto\mathbf{1}$, the choice of the basis in $LL$-squark sector has impact on the results. Matrices $V_{uL},V_{dL}$, which are undetermined in the SM, are partially observable due to the flavor violation in the squark loops \cite{Misiak:1997ei}.}, in which
\begin{equation}
V_{uL}=V_{CKM}, V_{dL}=\mathbf{1},\label{BRS}
\end{equation}
is used \cite{Buras:1997ij,Buras:2004qb}. This of course generates the flavor violation in the $M^2_{\tilde{u},LL}$, affecting the chargino diagrams in $K^+\rightarrow\pi^+\nu\bar{\nu}$, while the relevant gluino and neutralino couplings remain flavor diagonal. 

However, the $\mathbf{m}^2_{\tilde{Q}}$ matrix itself could contain off-diagonal components. We parametrize them separately in terms of $\Delta^{ij}_{\tilde{q}LL}$ as follows:
\begin{subequations}
\begin{eqnarray}
M^2_{\tilde{u},LL}&=&V_{CKM}\mathbf{m}^2_{\tilde{Q}}V^{\dagger}_{CKM}+\Delta_{\tilde{u}LL}
+\ldots,\\
M^2_{\tilde{d},LL}&=&\mathbf{m}^2_{\tilde{Q}}+\Delta_{\tilde{d}LL}+\ldots,
\end{eqnarray}
\end{subequations}
where $\ldots$ stands for other flavor diagonal terms (see Eq.~(\ref{M2sq})) and
\begin{equation}
\Delta_{\tilde{u}LL}=V_{CKM}\Delta_{\tilde{d}LL}V^{\dagger}_{CKM}.
\end{equation}
It is important to note that this parametrization can also be understood as deviation from Eq.~(\ref{BRS}), using Eq.~(\ref{M2sq}) with $V_{uL},V_{dL}$ composed of eigenvectors of
\begin{equation}
V_{CKM}\mathbf{m}^2_{\tilde{Q}}V^{\dagger}_{CKM}+\Delta_{\tilde{u}LL},
\mathbf{m}^2_{\tilde{Q}}+\Delta_{\tilde{d}LL},
\end{equation}
respectively. Therefore, the effect of flavor violation in the left-left squark sector can in principal be traced to quark Yukawa couplings in the electroweak basis as well as off-diagonal components of $\mathbf{m}^2_{\tilde{Q}}$.
In analogy to Eq.~(\ref{deltaNMFV}) we define
\begin{equation}\label{LLdeltaNMFV}
\delta^{ij}_{\tilde{q}LL}=\frac{(\Delta_{\tilde{q},LL})^{ij}}{\sqrt{(M^2_{\tilde{q},LL})^{ii}(M^2_{\tilde{q},LL})^{jj}}}.
\end{equation}
Assuming the soft scalar masses large enough compared to the $m_t$, we can take the denominators in Eq.~(\ref{LLdeltaNMFV}) to be approximately equal for $d$ and $u$ type squarks. Consequently,
\begin{equation}
\delta^{ij}_{\tilde{d}LL}=V^{\dagger}_{CKM}\delta^{ij}_{\tilde{u}LL}V_{CKM}.
\end{equation}

\subsection{Supersymmetry in \texorpdfstring{$K^+\rightarrow\pi^+\nu\bar{\nu}$}{Kaon} decay amplitude}

\begin{figure}[b]
\centerline{\includegraphics[width=10cm]{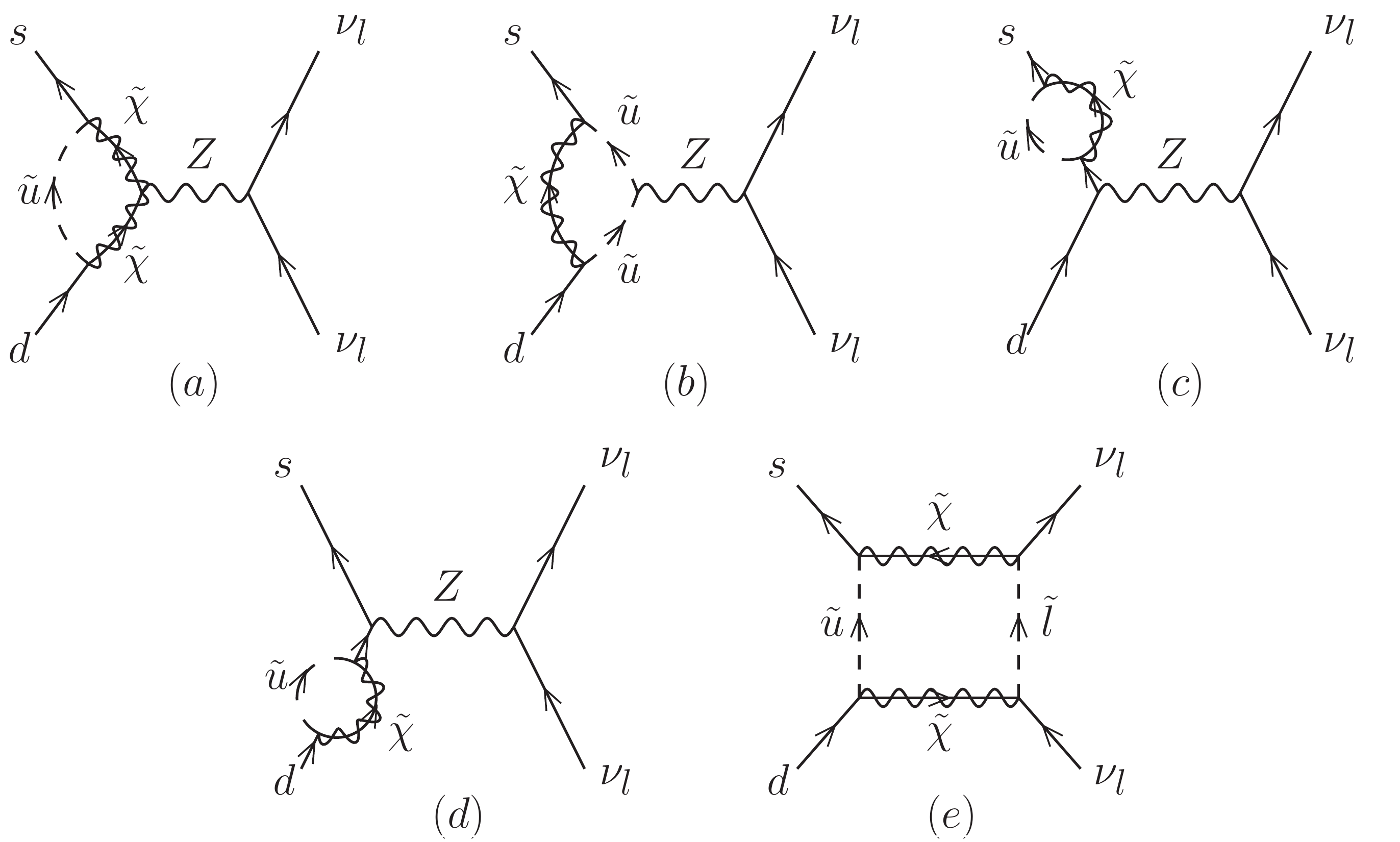}}
\caption{MSSM chargino diagrams for the $K^+\rightarrow\pi^+\nu\bar{\nu}$. 
\label{fig:Chargino diagrams}}
\end{figure}

flavor changing supersymmetric diagrams contributing to the $K^+\rightarrow\pi^+\nu\bar{\nu}$ decay amplitude contain four different sparticle combinations. Usually the most important effect comes from the chargino-up-squark loops in the $Z$-penguin diagrams (see Fig.~\ref{fig:Chargino diagrams}) \cite{Buras:1999da,Nir:1997tf,Colangelo:1998pm}. Neglecting the $s$ and $d$ quark masses, we obtain
\begin{eqnarray}
X^{\tilde{\chi}^\pm}_L=-\frac{1}{2}\frac{V^*_{is}V_{jd}}{V^*_{ts}V_{td}}\Big(R^{\tilde{u}*}_{\alpha i}V_{A1}-\frac{m_{u_i}}{\sqrt{2}M_W s_\beta}R^{\tilde{u}*}_{\alpha i+3}V_{A2}\Big)
\Big(R^{\tilde{u}}_{\beta j}V^{\tilde{\chi}*}_{B1}\\-\frac{m_{u_j}}{\sqrt{2}M_W s_\beta}R^{\tilde{u}}_{\beta j+3}V^{\tilde{\chi}*}_{B2}\Big)
\Big(Y^{(a)}_{\alpha\beta AB}+Y^{(b)}_{\alpha\beta AB}+Y^{(c+d)}_{\alpha\beta AB}+Y^{(e)}_{\alpha\beta AB}\Big),\nonumber
\end{eqnarray}
where
\begin{eqnarray}
Y^{(a)}_{\alpha\beta AB} =
-\delta_{AB}\big(\tfrac{4}{3}s^2_W\delta_{\alpha\beta}-R^{\tilde{u}}_{\alpha k}R^{\tilde{u}*}_{\beta k}\big)  C_{00}(M_{\tilde{\chi}_A},M_{\tilde{u}_\alpha},M_{\tilde{u}_\beta}),
\label{Yns1}\\
Y^{(b)}_{\alpha\beta AB} =
\delta_{\alpha\beta}\big(\big(\tfrac{1}{2}-s^2_W\big)\delta_{AB}+\tfrac{1}{2}U^{\tilde{\chi}*}_{A1} U_{B1}\big) M_{\tilde{\chi}_A}M_{\tilde{\chi}_B}C_0(M_{\tilde{\chi}_A},M_{\tilde{\chi}_B},
M_{\tilde{u}_\alpha})
\label{Yns2}\\
-2\delta_{\alpha\beta}\big(\big(\tfrac{1}{2}-s^2_W\big)\delta_{AB}+\tfrac{1}{2}V_{A1}V^{\tilde{\chi}*}_{B1}\big)\nonumber C_{00}(M_{\tilde{\chi}_A},M_{\tilde{\chi}_B},M_{\tilde{u}_\alpha}),\\
Y^{(c+d)}_{\alpha\beta AB} =
-\delta_{\alpha\beta}\delta_{AB}\big(\tfrac{1}{2}-\tfrac{1}{3}s^2_W\big) B_1(M_{\tilde{\chi}_A},M_{\tilde{u}_\alpha}),
\label{Yns3}\\
Y^{(e)}_{\alpha\beta AB} =
-U_{A1}U^{\tilde{\chi*}}_{B1}\delta_{\alpha\beta}
R^{\tilde{l}}_{\gamma l}R^{\tilde{l}*}_{\gamma l}M_{\tilde{\chi}_A}M_{\tilde{\chi}_B}  D_0(M_{\tilde{\chi}_A},M_{\tilde{\chi}_B},M_{\tilde{u}_\alpha},M_{\tilde{l}_\gamma}).
\label{Yns4}
\end{eqnarray}
Loop functions we use, $B_0$, $B_1$, $C_0$, $C_{00}$ and $D_0$, are defined in Ref. \citen{Hahn:1998yk}. One can easily observe that the VEV-independent parts of Eq.~(\ref{Yns1})-(\ref{Yns4}) (those proportional $\delta_{\alpha\beta}\delta_{AB}$) completely cancel out and, even though the size of $Y^{(a)}$ and $Y^{(b)}$ is of the same order of magnitude as the dominant SM part, overall chargino contribution to $X_L$ is reduced by two orders of magnitude. The reason is that exact $SU(2)_L$ invariance would not allow for generated $s-d-Z$ effective coupling and necessary symmetry breaking is needed \cite{Colangelo:1998pm,Buras:1997ij}. The same applies to $X^{\tilde{\chi}^\pm}_R$, which is even more suppressed due to presence of $m_s m_d/M^2_W$ coming from $\tilde{\chi}_A-\tilde{u}_\alpha-d,s$ vertices. 

There is potentially large effect coming from the charged Higgs loops. Its couplings are proportional to the fermion masses and therefore untouched by the cancellation mentioned above, but these diagrams including $X_R\propto\tan^2\beta$ are also suppressed by smallness of $m_s, m_d$. Furthermore, the charged Higgs loops depend on $m^2_t/M^2_{H^\pm}$ being small for reasonable heavy charged Higgs. Therefore, overall supersymmetric effects in the MFV case are very limited. However, the charged Higgs loops become important if $\tan\beta$ is large and right-right block of the down squark mass matrix contains non-minimal flavor violation \cite{Isidori:2006jh}.

Gluino and neutralino-down-squark loops arise in the case of non-minimal flavor violation (NMFV), only. Their role is limited by the fact that either double chirality flips on squark propagators, or left-right non-minimal flavor mixing terms, lead to the factors of $m^2_{d_i}$ over some large mass squared \cite{Buras:1997ij}.

\section{NMFV and left-left mixing in \texorpdfstring{$K^+\rightarrow\pi^+\nu\bar{\nu}$}{charged kaon decay}}

The effect of $\delta^{ij}_{\tilde{q}XY}$ on FCNC amplitude can be large and many of these mass insertions are already constrained by observed absence of a deviation from the standard model predictions. In the literature, the following two types of NMFV relevant to the $K^+\rightarrow\pi^+\nu\bar{\nu}$ decay are mentioned:
\begin{romanlist}[(ii)]
\item $\delta_{\tilde{u}LR}$ - the left-right mass insertions provide necessary electroweak symmetry breaking and no other $v_{u,d}$ are needed. Sensitivity of the $K^+\rightarrow\pi^+\nu\bar{\nu}$ to the $\delta^{13,23}_{\tilde{u}LR}$ is well known and enhanced for a small values of $\tan\beta$ \cite{Buras:1997ij,Buras:2004qb,Nir:1997tf,Isidori:2006qy}. Sizes of $\delta_{\tilde{u}LR}$ are constrained by the charge and color symmetry breaking condition (CCB) as well as the fact that the scalar potential should be bounded from bellow (UFB). Both this bounds can be written as
\begin{align*}
(M^2_{\tilde{u},LR})^{ij}<m_{u_k}\sqrt{m^2_{\tilde{Q}_i}+m^2_{\tilde{u}_j}+
\min\{m^2_{H_u},m^2_{\tilde{L}_i}+m^2_{\tilde{l}_j}\}},
\end{align*}
where $i\neq j$ and $k=\max\{i,j\}$ \cite{Jager:2008fc,Casas:1996de}.
\item $\delta_{\tilde{d}RR}$ - the suppression of $X_R$ by small down quark masses can be overcome by non-minimal flavor violation in the right-right $d$-squark sector, giving arise $H^\pm-u_i-d,s$ effective vertices originating predominantly from the $\tilde{g}-\tilde{u}_L-\tilde{d}_R$ loops \cite{Isidori:2006jh}. Leading terms, proportional to the $\delta^{13}_{\tilde{d}RR}\delta^{23}_{\tilde{d}RR}$ and $\delta^{13}_{\tilde{d}RR}$, scale approximately as $\tan^4\beta$ and $\tan^3\beta$, respectively. They are not suppressed by small CKM elements and in spite of the factor of $m^2_b/M^2_W$ can allow for effects as large as tenths of percent depending on the charged Higgs and gluino masses \cite{Isidori:2006jh}. 
\end{romanlist}

In this work we point out that third class of mass insertions, leading to not very large, but still measurable effect in the decay branching ratio, has to be taken into account. In previous analyses it has been often noted that the left-left flavor breaking terms, even of potential importance in rare kaon decays, are strictly constrained by the measurements of $\Delta F=2$ processes, such as $\Delta M_K,\epsilon_K$, $\Delta M_{d,s}$ \cite{Colangelo:1998pm,Buras:2004qb,Isidori:2006qy}, and their contribution to our decay is strongly suppressed \cite{Colangelo:1998pm}. Of course, both this statements were true. However, there are at least two good reasons, why we should reinvestigate the effect of the left-left squark mixing on $K^+\rightarrow\pi^+\nu\bar{\nu}$, taking the recent experimental results into account. Within the GUT motivated MSSM, we expect heavy glui\-no masses. This fact is well justified by recent chargino and neutralino searches by CMS \cite{Chatrchyan:2012pka,Chatrchyan:2013sza} and ATLAS \cite{ATLAS-CONF-2013-035}, both leading to similar lower bounds for bino-like neutralino mass\-es, $M_1>400$ GeV. Then, as a consequence of Eq.~(\ref{GUT}), we get $M_2>0.8$ TeV as well as $M_3>2.8$ TeV. 
\begin{table}[ph]
\tbl{Limits on left-left mass insertions for squark and gluino masses given in TeV.}
{\begin{tabular}{@{}lllrr@{}} \toprule
$\delta^{ij}_{\tilde{q}LL}$ & constraining observables & upper bound & $\tilde{m}$ & $M_3$\\ \colrule
$|\delta^{12}_{\tilde{u}LL}|$ & $D^0-\bar{D}^0$ & $0.10$ \cite{Ciuchini:2007cw,Gedalia:2009kh} & $<1.0$ & $<1.0$\\ 
& & $0.14$ \cite{Ciuchini:2007cw} & $0.5$ & $1.0$\\ 
& & $0.06$ \cite{Altmannshofer:2009ne,Ciuchini:2007cw,Jager:2008fc} & $< 0.6$ & $<0.6$\\ \colrule
$|\delta^{12}_{\tilde{d}LL}|$ & $K^0-\bar{K}^0$ & $0.14$ \cite{Buras:2004qb} & $<1.0$ & $<2.0$\\
$|\mathrm{Re}(\delta^{12}_{\tilde{d}LL})|$ & & $0.03$ \cite{Altmannshofer:2009ne,Jager:2008fc} & $< 0.6$ & $<0.6$\\
$|\mathrm{Im}(\delta^{12}_{\tilde{d}LL})|$ & & $0.003$ \cite{Jager:2008fc} & $0.5$ & $0.5$\\ \colrule
$|\mathrm{Re}(\delta^{13}_{\tilde{d}LL})|$ & $\Delta M_d$, $S_{\psi K_S}$ & $0.1$ \cite{Altmannshofer:2009ne,Jager:2008fc} & $<0.6$ & $<0.6$\\ 
$|\mathrm{Im}(\delta^{13}_{\tilde{d}LL})|$ & & $0.03$ \cite{Altmannshofer:2009ne} & $<0.6$ & $< 0.6$\\ \colrule
 & & & $\tilde{m}$ & $M_2$\\ \colrule
$|\delta^{13}_{\tilde{d}LL}|$ & $B\rightarrow$ $X_s\gamma$, $X_s l\bar{l}$ & $0.24$ \cite{Xiao:2006gu} & $0.5$ & $0.6$ \\
$|\delta^{23}_{\tilde{d}LL}|$ & & $0.11$ \cite{Xiao:2006gu} & $0.5$ & $0.6$ \\
$|\mathrm{Re}(\delta^{23}_{\tilde{d}LL})|$ & & $0.1$ \cite{Altmannshofer:2009ne} & $<0.6$ & $< 0.16$\\
$|\mathrm{Im}(\delta^{23}_{\tilde{d}LL})|$ & & $0.2$ \cite{Altmannshofer:2009ne} & $<0.6$ & $<0.16$\\ \botrule
\end{tabular}\label{NMFVtable}}
\end{table}
Since the largest supersymmetric contribution to the $\Delta F = 2$ observables comes from the gluino mediated NMFV diagrams, the left-left squark mixing constrains will be significantly weaker compared to those listed in the Table~\ref{NMFVtable}. We also consider heavy pseudoscalar Higgs, $m_{A^0}\simeq 1-2$ TeV and heavier squarks with the masses of the order of $1$ TeV. Both these assumptions, motivated by measurements of the $B^0_{s}\rightarrow\mu^+\mu^-$ decay, weaken the limits on the left-left mixing.

\begin{figure}[b]
\centerline{\includegraphics[width=8cm]{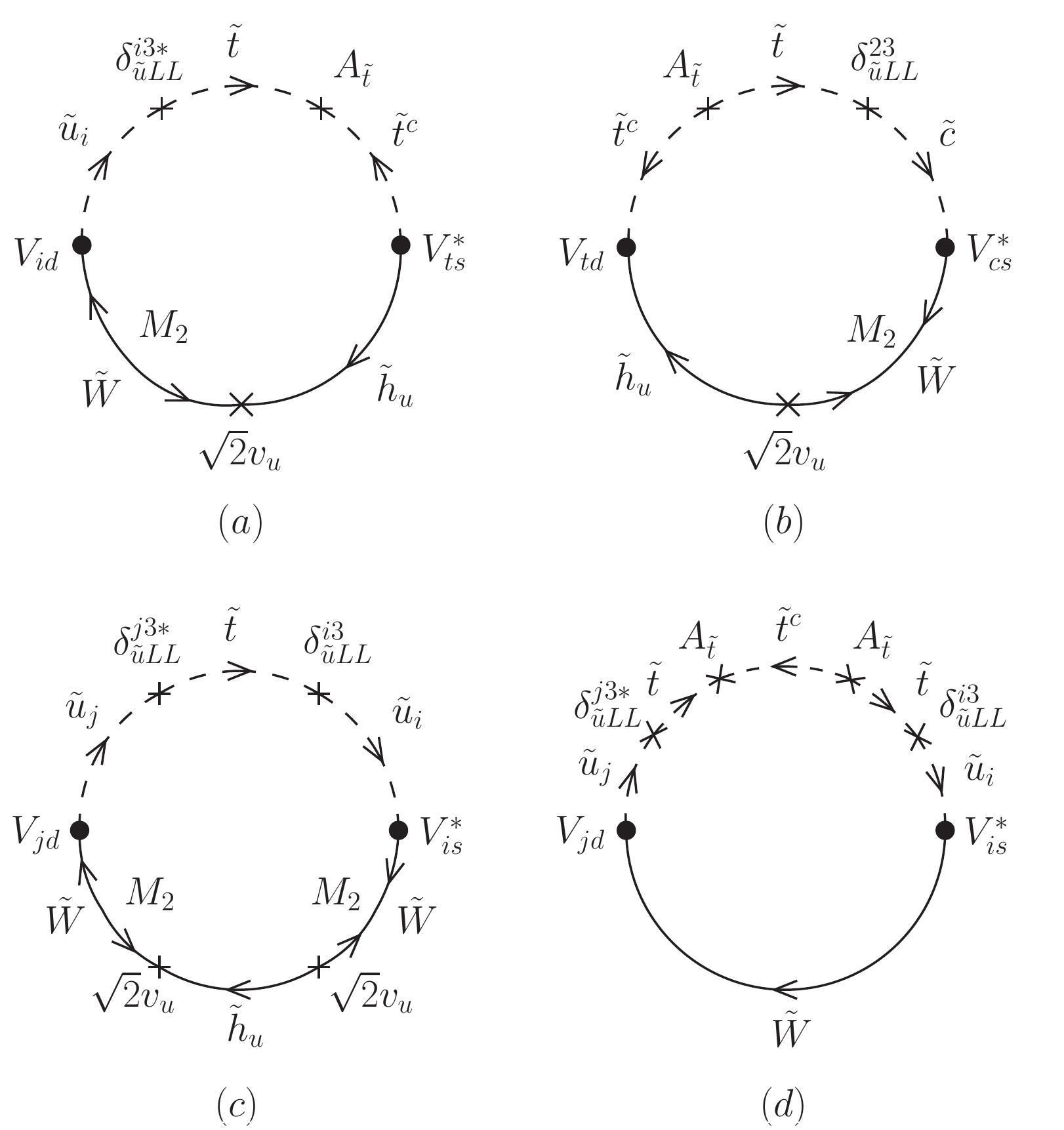}}
\caption{Dominant $\delta^{i3}_{\tilde{u}LL}$ loops contributing to $K^+\rightarrow\pi^+\nu\bar{\nu}$ decay in the large $\tan\beta$ regime. $\tilde{W}$ and $\tilde{h}_u$ are represented by two-component Weyl spinors.
\label{fig:MIAloops}}
\end{figure}

The second mechanism increasing the importance of $\delta_{\tilde{u}LL}$ comes from the measurement of Higgs mass, equal $125.03^{+0.26}_{-0.27}\mathrm{(stat.)}^{+0.13}_{-0.15}\mathrm{(syst.)}$ GeV\cite{CMS-PAS-HIG-14-009}. It is a well known fact that in supersymmetry the Higgs mass, at the tree level obeying the condition $m_{h^0}<M_Z |\cos2\beta|$, receives important corrections dominated by the top and stop loops. Hence, in order to obtain observed value of $m_{h^0}$ one needs to employ large $\vert A_{\tilde{t}} \vert$, amounting to $2 - 4$ TeV (see Fig.~\ref{fig:Higgs}). Again, this was not the case of the studies leading to limits in the Table~\ref{NMFVtable}, since the Higgs mass was unknown and values of $|A_{\tilde{t}}|\simeq 0.1-1$ TeV were used \cite{Buras:2004qb,Nir:1997tf, Isidori:2006qy}. As already mentioned, the size of left-right scalar mixing is crucial for the $K^+\rightarrow\pi^+\nu\bar{\nu}$ amplitude. Therefore, we came to the conclusion that similar mechanism making the effect of $\delta^{13,23}_{\tilde{u}LR}$ dominant in $K^+\rightarrow\pi^+\nu\bar{\nu}$ decay, can be applied (at least to some extent) to the left-left flavor mixing as well. In this case, the double insertion of $A_{\tilde{t}}$ and $\delta^{13,23}_{\tilde{u}LL}$ into chargino-up-squark loop share the role of $SU(2)_L$ and flavor violation, respectively. 

The clearest way to demonstrate the effect of supersymmetry in flavor violating decays consist of drawing the loops in the mass insertion approximation. Our aim here is to understand the role of different flavor and gaugino mixing parameters. Therefore, we treated in this approximation wino and Higgsino as well, both represented by two-component Weyl spinors. This means that we have, for large $\tan\beta$, either Higgsino and wino propagators mixed via $\sqrt{2}v_u$, or one wino propagator with extra $M_2$. The $\tilde{h}_u-\tilde{h}_d$ Higgsino propagator containing extra $\mu$ insertion is always suppressed by $\cos\beta$ and can be neglected. 

The contributions corresponding to $\delta^{i3}_{\tilde{u}LL}$ are depicted in Fig.~\ref{fig:MIAloops}. Each of $(a)-(d)$ represents sum of all possible penguin diagrams, in which the loop may be contained. That includes vertex corrections with $Z$-propagator connected to all the squark, Higgsino or wino internal legs as well as diagrams with the loop on external leg. The dependence of the decay amplitude on $12$ left-left delta can be well approximated by the loop similar to $(c)$, with single $\delta^{12}_{\tilde{u}LL}$ insertion instead of $\delta^{i3}_{\tilde{u}LL} \delta^{j3*}_{\tilde{u}LL}$. One can observe that diagrams $(a),(b)$ and $(d)$ perfectly fit the pattern in which double insertion of $\delta^{i3}_{\tilde{u}LL}$ and $A_{\tilde{t}}$ plays the role of $\delta^{i3}_{\tilde{u}LR}$. It is also easy to see that the contribution of $\delta^{13}_{\tilde{u}LL}$ will be dominated by the loop $(a)$ only, as the diagram similar to $(b)$ would be suppressed by very small CKM factor. Loops $(c)$ and $(d)$ contain one more insertion, which, in general, makes them smaller compared to $(a),(b)$. Moreover, they have opposite sign to each other and for certain values of $A_{\tilde{t}}$ and $M_2$ can completely cancel. 

\begin{table}[ph]
\tbl{Sensitivity of $(a)-(d)$ chargino-up-squark loops from Fig.~\ref{fig:MIAloops} on $\delta^{13,23}_{\tilde{u}LL}$.}
{\begin{tabular}{@{}rrrrr@{}} \toprule
& $X^{(a)}_L/X^{SM}_L$ & $X^{(b)}_L/X^{SM}_L$ & $X^{(c)}_L/X^{SM}_L$ & $X^{,(d)}_L/X^{SM}_L$\\ \colrule
$\delta^{13}_{\tilde{u}LL}=0.3$ & $0.12$ & $-0.01$ & $-0.14$ & $0.07$\\
$\delta^{23}_{\tilde{u}LL}=0.3$ & $-0.03$ & $-0.03$ & $0.14$ & $-0.07$\\ \botrule
\end{tabular}\label{MIAtable}}
\end{table}

Relative importance of the corresponding diagrams can be seen in the Table~\ref{MIAtable}, where $m_{\tilde{Q}_3}=1.2$ TeV, $A_{\tilde{t}}=-2.2$ TeV and values from Table~\ref{Paramtable} have been used. Both plots show the contribution of non-minimal flavor violation to the $X_L$ coming from chargino-up-squark diagrams. The full expressions for all the loops in Fig.~\ref{fig:MIAloops} can be found in appendix \ref{A}.

\section{Numerical Results}

\begin{figure}[b]
\centerline{\includegraphics[width = 8cm]{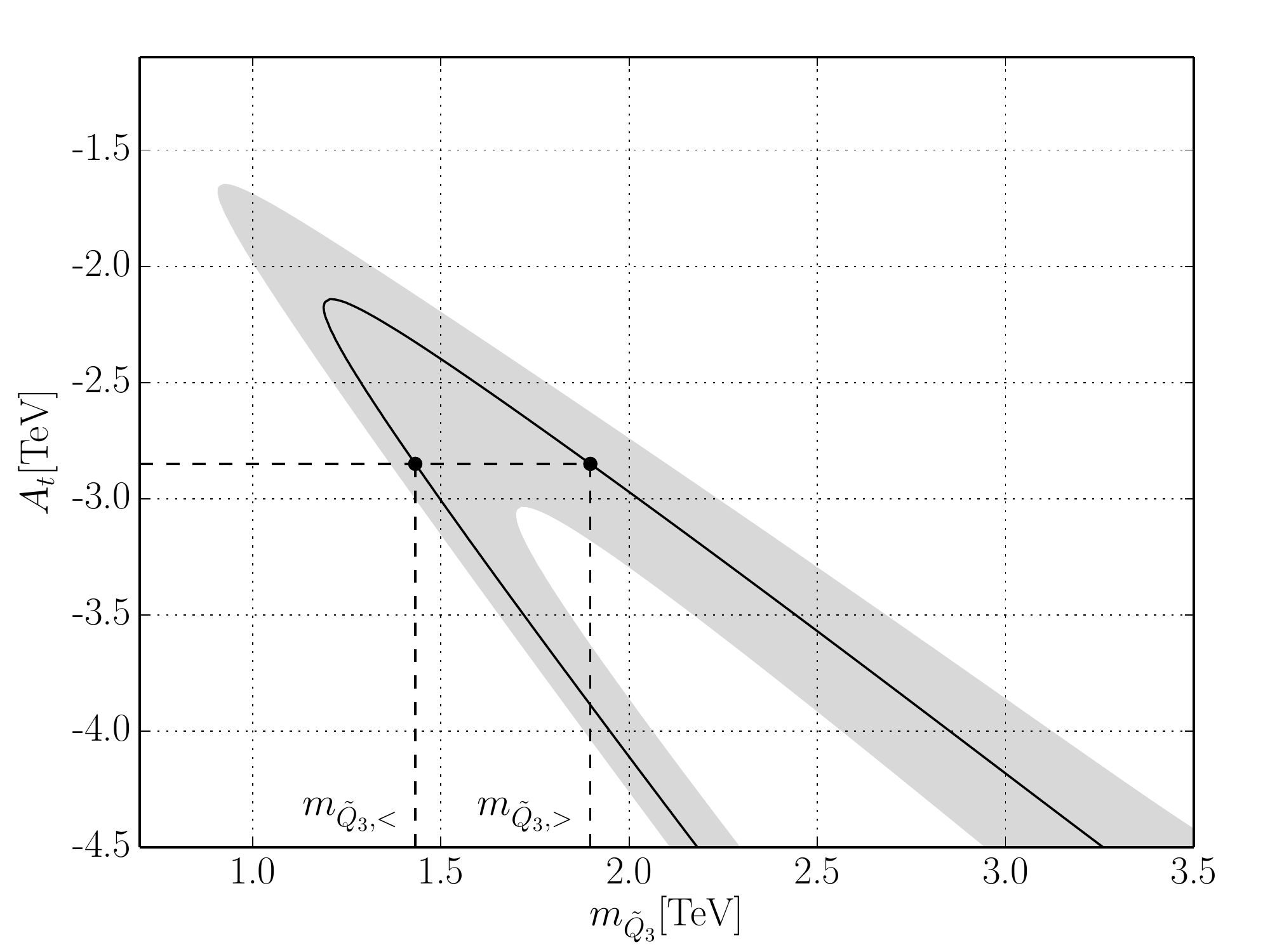}}
\caption{Allowed values of the scalar mass and left-right mixing corresponding to $\tan\beta = 50$ and Higgs mass measured by CMS. Solid line pertains to the mean value, while the shaded pink region shows how it varies when one sigma deviations of $m_{h^0} = 125.03^{+0.26}_{-0.27} \mathrm{(stat.)}^{+0.13}_{-0.15} \mathrm{(syst.)}$ GeV \cite{CMS-PAS-HIG-14-009} and $M_t = 173.34\pm 0.27 (\mathrm{stat.})\pm 0.71 (\mathrm{syst.})$ GeV \cite{ATLAS:2014wva} are taken into account. Two different stop masses denoted $m_{\tilde{Q}_3,>}$, $m_{\tilde{Q}_3,<}$ are possible for each value of $A_{\tilde{t}}$.
\label{fig:Higgs}}
\end{figure}

For our numerical analysis of the flavor changing processes we have used the fortran library {\tt SUSY\_FLAVOR} v2.11 \cite{Crivellin:2012jv}. As already mentioned in the previous section, the recent measurement of Higgs mass leads to constraint on the left-right up-squark mixing and the stop mass, directly related to the quantities $A_{\tilde{t}}$ and $m_{\tilde{Q}_3}$, respectively. Therefore, we first used FeynHiggs 2.10.1 \cite{Hahn:2013ria, Frank:2006yh, Degrassi:2002fi, Heinemeyer:1998np, Heinemeyer:1998yj} to calculate Higgs mass. Then, $m_{h^0}$ as a function of supersymmetric parameters sensitive mostly to $A_{\tilde{t}}, m_{\tilde{Q}_3}, m_{\tilde{u}_3}$ and $\tan\beta$, was numerically inverted. Fixing the value of the $\tan\beta$ equal $50$ allows for two values of $m_{\tilde{Q}_3}$ related to each $A_{\tilde{t}}$, named as shown in Fig.~\ref{fig:Higgs}. 

We assumed specific values of the MSSM parameters that were different from the often used, though very restricted, CMSSM. Taking gauge unification into account, as a consequence of large third family Yukawa couplings in the RGE, we obtain $m_{\tilde{Q}_1}>m_{\tilde{Q}_3}$. Such an assumption, in principle, generates flavor off-diagonal terms in the $M^2_{\tilde{u}LL}$ given by Eq.~(\ref{MFVblocks1})-(\ref{MFVblocks4}). However, this effect turned out to be negligible in $K^+\rightarrow\pi^+\nu\bar{\nu}$ due to the small CKM matrix elements. 

\begin{table}[h]
\tbl{Used values of the MSSM parameters. All masses are in TeV.}
{\begin{tabular}{@{}ccccccc@{}} \toprule
$M_2$ & $\mu$ & $M_{A^0}$ & $\tan\beta$ & $m_{\tilde{Q}_1}$ & $m_{\tilde{u}_1,\tilde{d}_1}$ & $m_{\tilde{u}_3,\tilde{d}_3}$\\ \colrule
$1.0$ & $0.11$ & $1.5$ & $50$ & $1.3\times m_{\tilde{Q}_3}$ & $1.3\times m_{\tilde{Q}_3}$ & $m_{\tilde{Q}_3}$\\ \botrule
\end{tabular}\label{Paramtable}}
\end{table}

Values of all the necessary supersymmetric parameters are summarized in the Table~\ref{Paramtable}, where we omitted the soft slepton masses. They are assumed to be of the order $\mathcal{O}(1\mathrm{TeV})$ and have little impact on the decay branching ratio. For the gauge sector, once we fix the value of $M_2$, the bino and gluino masses are given by Eq.~(\ref{GUT}). The $A_{\tilde{t}}$, as well as the corresponding scalar mass obtained in terms of $m_{\tilde{Q}_3,>}$ or $m_{\tilde{Q}_3,<}$, is in our work taken as a free quantity and varied together with $\delta^{ij}_{\tilde{u}LL}$. However, in order to fulfill $B\rightarrow X_s \gamma$ experimental limits mentioned bellow, we have chosen negative sign of $A_{\tilde{t}}$.

\begin{figure}[b]
\centerline{
\subfloat{\includegraphics[width = 6.5cm]{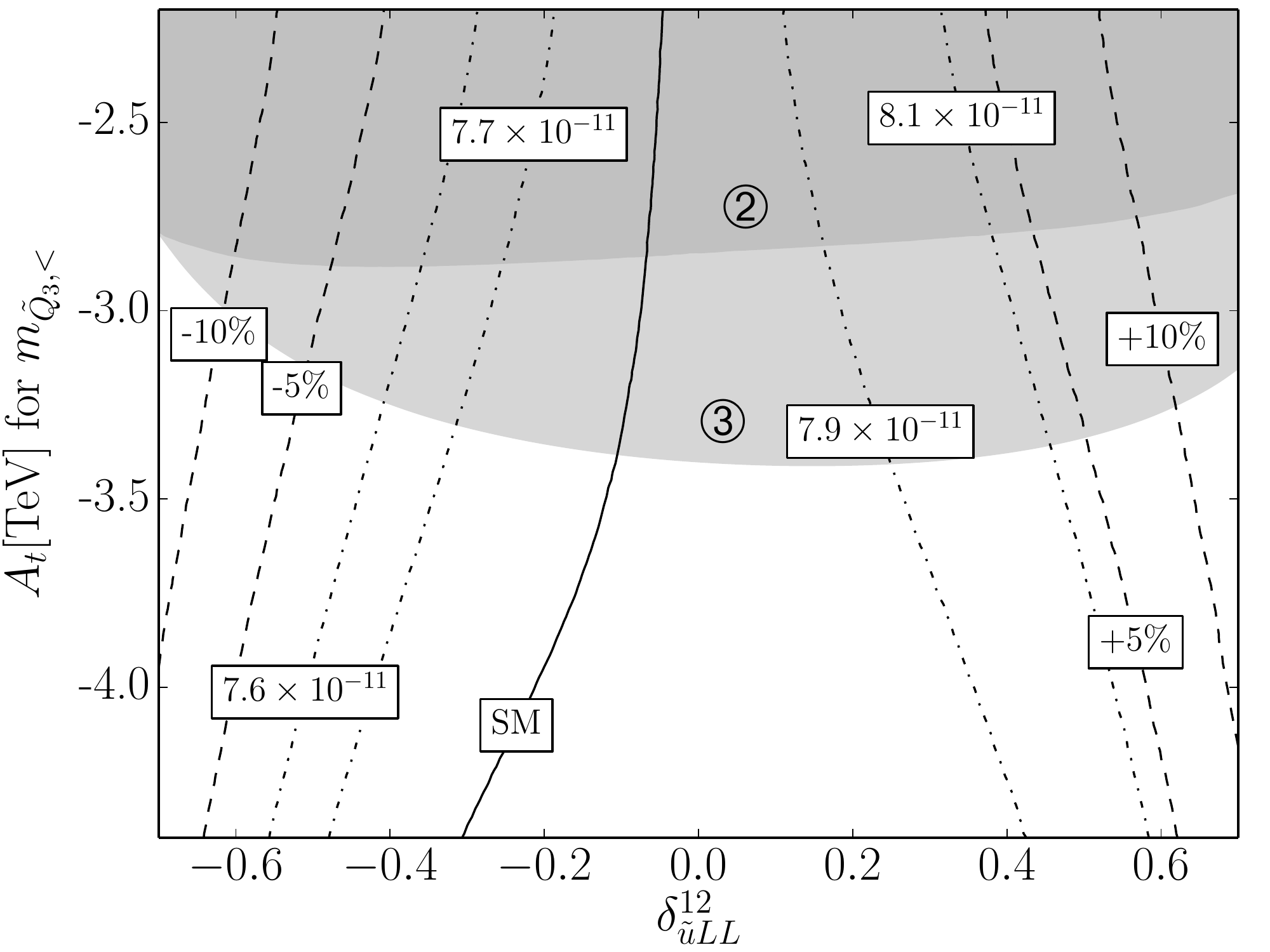}}~
\subfloat{\includegraphics[width = 6.5cm]{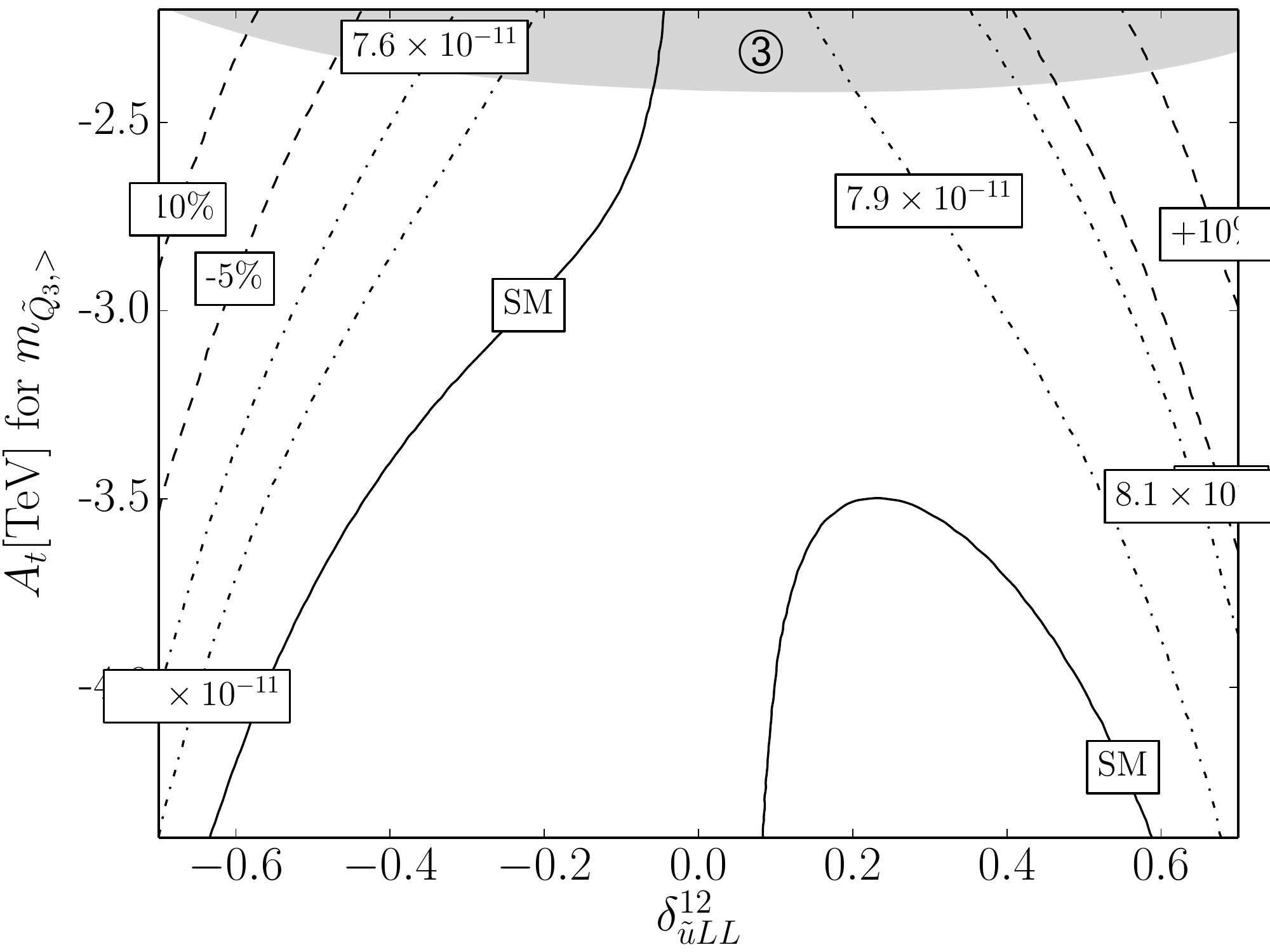}}}
\centerline{
\subfloat{\includegraphics[width = 6.5cm]{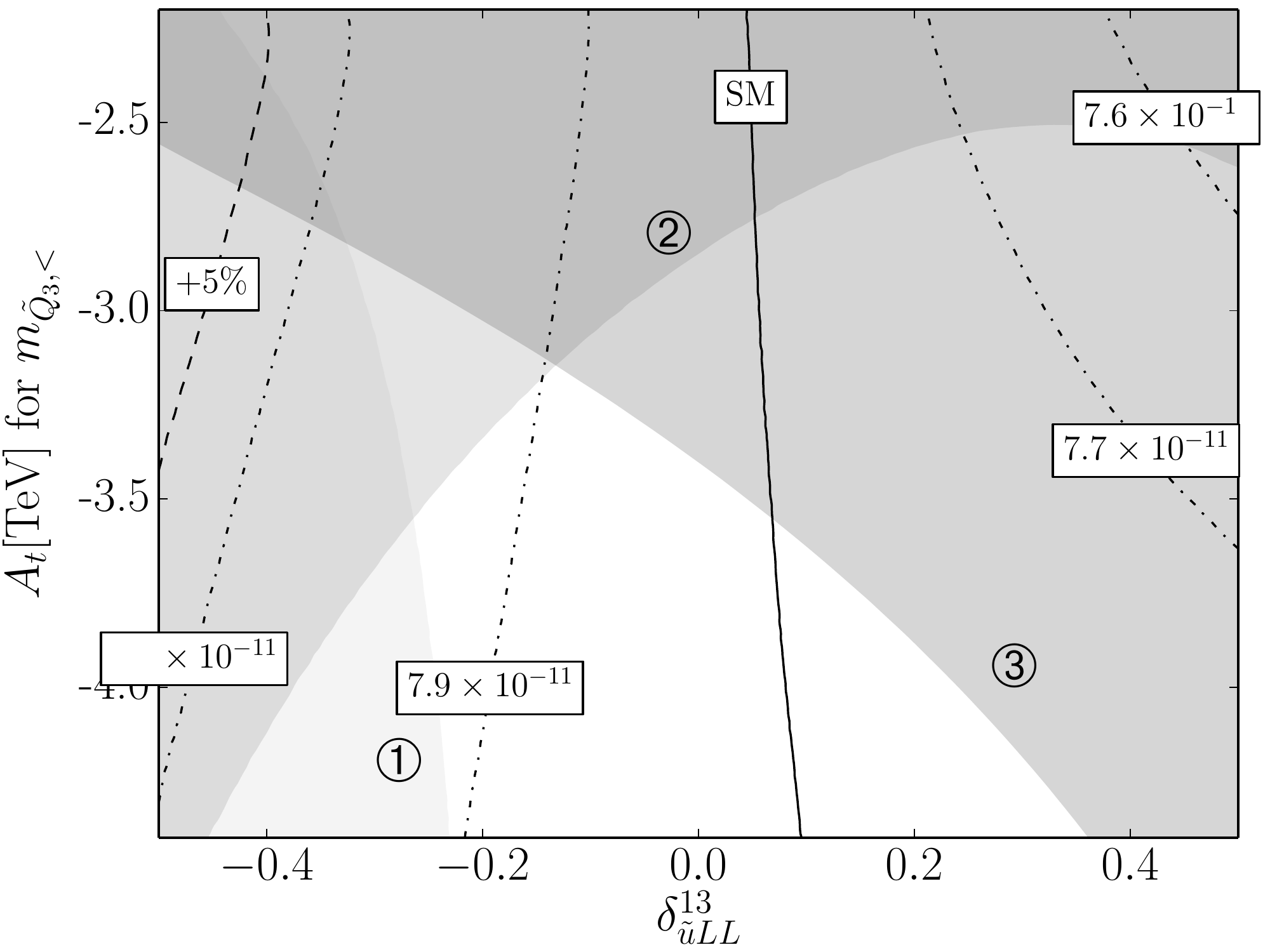}}~
\subfloat{\includegraphics[width = 6.5cm]{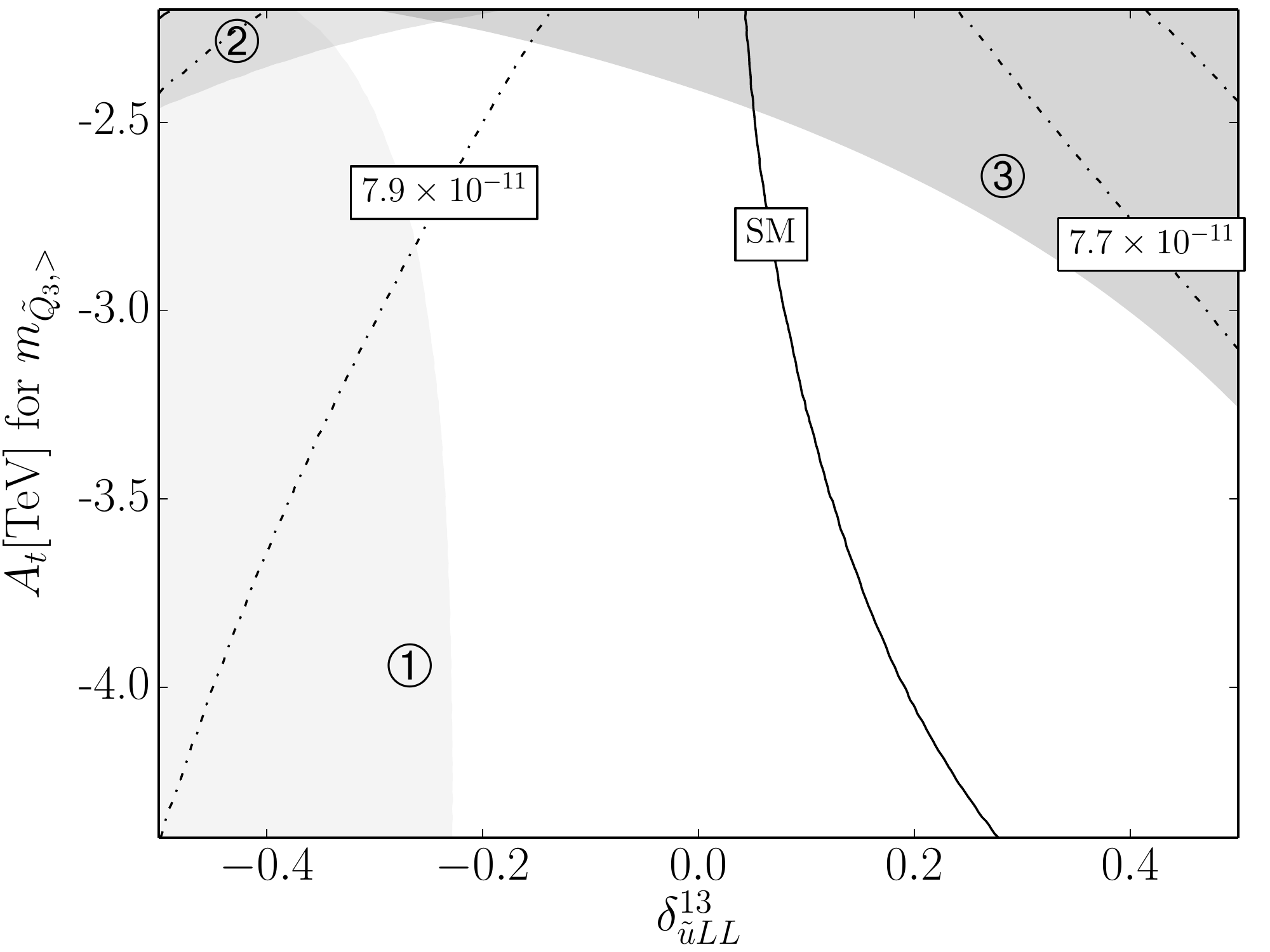}}}
\caption{The effect of $A_{\tilde{t}}$ and $\delta_{\tilde{u}LL}$ in the 		  $K^+\rightarrow\pi^+\nu\bar{\nu}$ decay branching ratio.\label{fig:uLL}}
\end{figure}

\begin{figure}[b]
\centerline{
\subfloat{\includegraphics[width = 6.5cm]{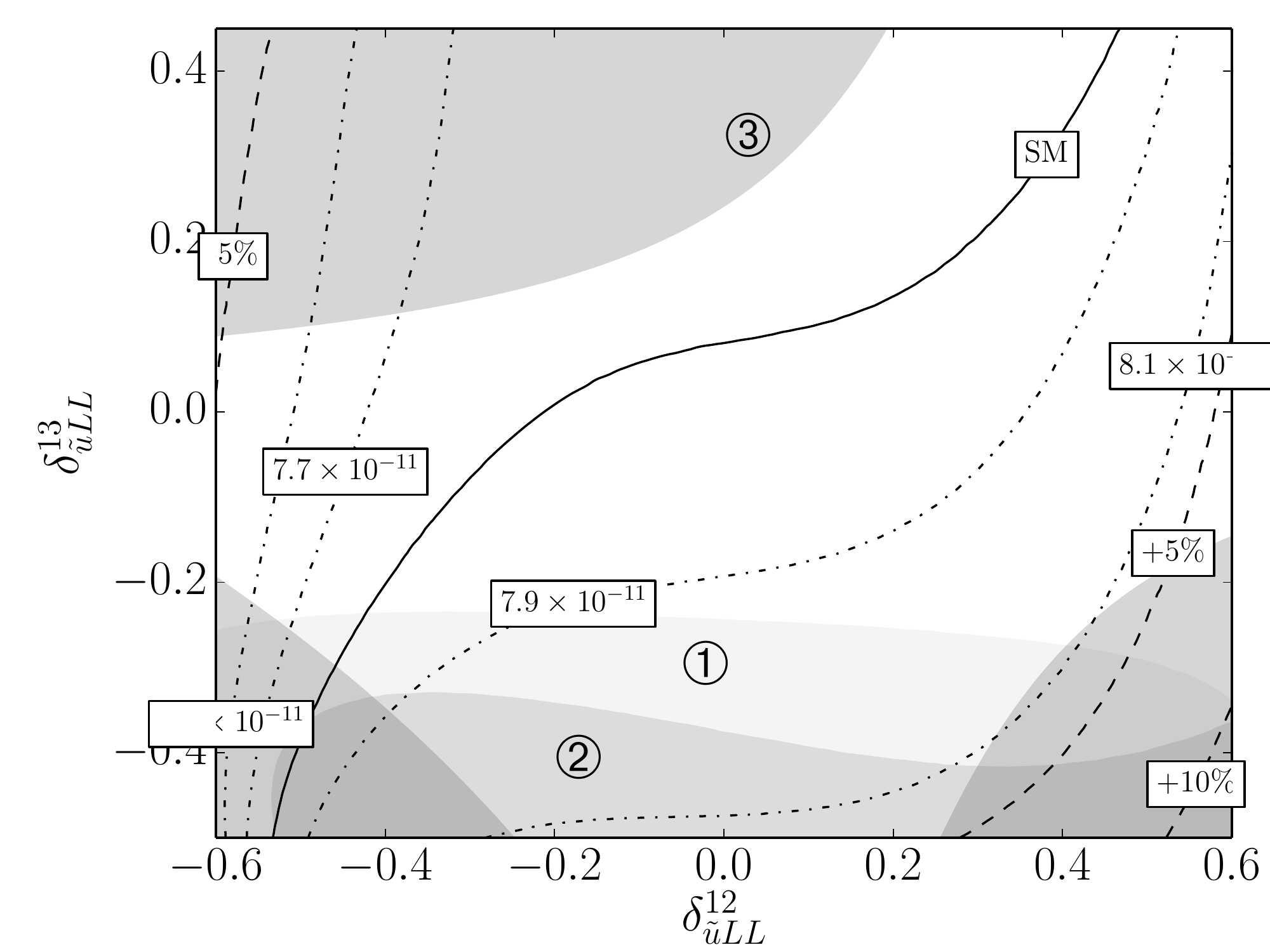}}~
\subfloat{\includegraphics[width = 6.5cm]{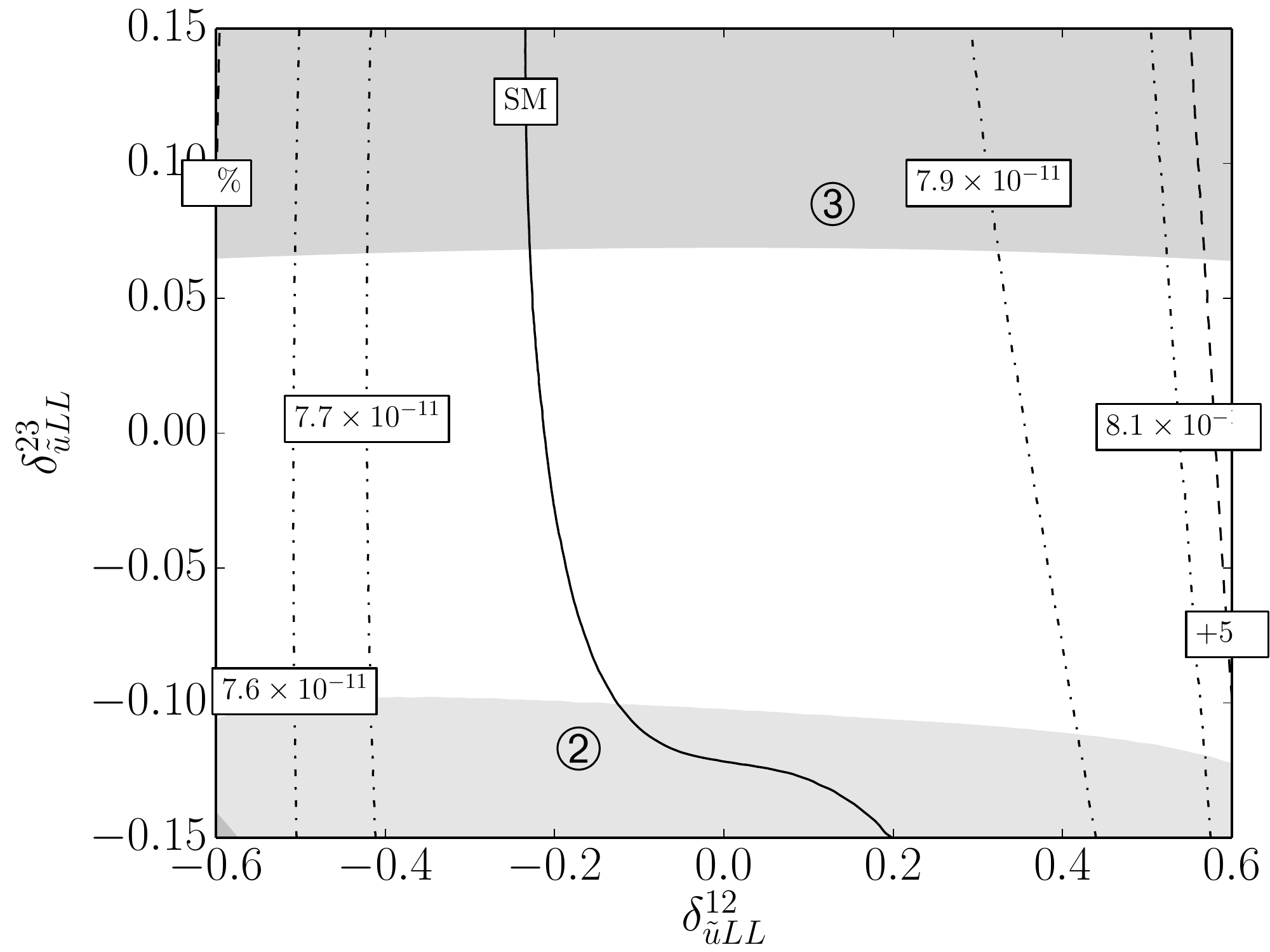}}}
\centerline{
\subfloat{\includegraphics[width = 6.5cm]{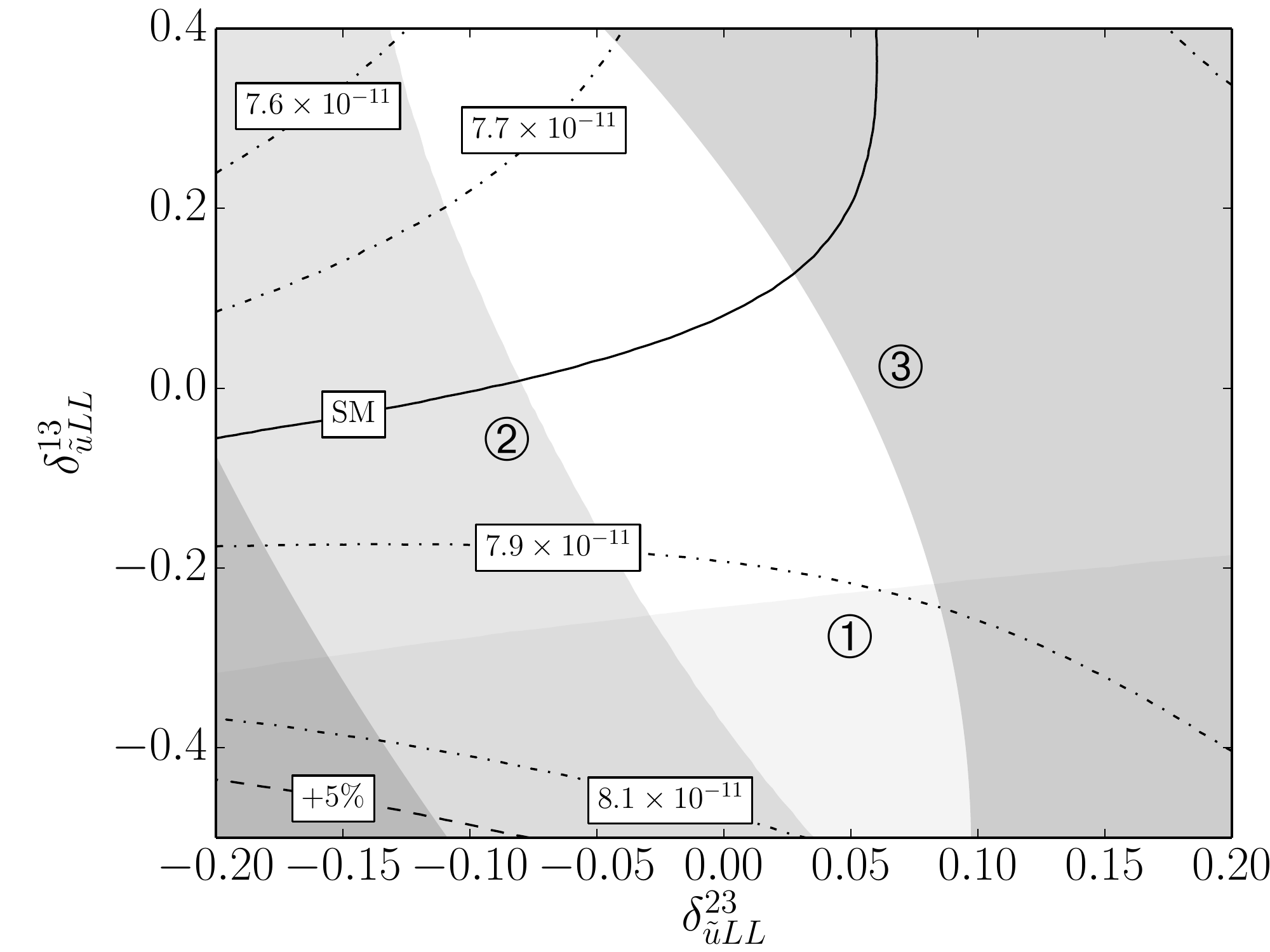}}}
\caption{Sensitivity of $K^+\rightarrow\pi^+\nu\bar{\nu}$ branching ratio to $\delta^{12}_{\tilde{u}LL},\delta^{13}_{\tilde{u}LL}$, $\delta^{12}_{\tilde{u}LL},\delta^{23}_{\tilde{u}LL}$ and $\delta^{13}_{\tilde{u}LL},\delta^{23}_{\tilde{u}LL}$ insertions, for $m_{\tilde{Q}_3},A_{\tilde{t}}$ fixed at $1.95$ TeV and $-4.0$ TeV, respectively.}
\label{fig:uLLuLL}
\end{figure}

The dependence of $BR(K^+\rightarrow\pi^+\nu\bar{\nu})$ on $A_{\tilde{t}}$ and $\delta^{12,13}_{\tilde{u}LL}$\footnote{We do not show the dependence on the $\delta^{23}_{\tilde{u}LL}$ as it has almost no effect on the $K^+\rightarrow\pi^+\nu\bar{\nu}$ branching ratio and its values are strongly constrained by the $B\rightarrow X_s\gamma$ measurement as well.} is shown in Fig.~\ref{fig:uLL}. All the flavor braking insertions are considered to be real. The solid line represents the values reproducing the standard model prediction, while the dashed contours show the size of the corresponding insertion for which $\pm5\%$ or $\pm10\%$ effect is reached. For each $\delta^{ij}_{\tilde{u}LL}$ there are two different plots, as we have two possible scalar soft masses corresponding to each value of $A_{\tilde{t}}$. In this figure, as well as in Fig.~\ref{fig:uLLuLL}, the gray shaded regions correspond to the values excluded at $95\%$ confidence level by \ding{192} $BR(B_d \rightarrow \mu^+\mu^-) = (3.6^{+1.6}_{-1.4})\times 10^{-10}$ \cite{CMSandLHCbCollaborations:2013pla}, \ding{193} $BR(B_s \rightarrow \mu^+\mu^-) = (3.6 \pm 2.9)\times 10^{-9}$ \cite{CMSandLHCbCollaborations:2013pla} and \ding{194} $BR(B \rightarrow X_s \gamma) = (3.31 \pm 0.35)\times 10^{-4}$ \cite{Lees:2012ym}. Parametric uncertainties coming from the top mass \cite{ATLAS:2014wva}, hadronic formfactors \cite{Davies:2012qf} and CKM matrix elements \cite{PhysRevD.86.010001} were added to the given experimental error of these processes. Resulting parametric errors read $2.5\%$, $8.5\%$ and $7.0\%$ in \ding{192}, \ding{193} and \ding{194}, respectively.

As one can expect, the effect of the insertion is more significant for the smaller of the two possible scalar masses, $m_{\tilde{Q}_3,<}$, which makes the loop functions less suppressed. However, we can see that in both cases the suppression due to increasing squark masses wins against the enhancement by the large scalar left-right mixing represented by $A_{\tilde{t}}$. Still, measurable effects up to $10\%$, consistent with the \ding{192} - \ding{194} constraints, are possible for $12$ left-left insertions.

The dependence of the kaon decay branching ratio in the case of two non-zero mass insertions is shown in the figure \ref{fig:uLLuLL}. The value of $A_{\tilde{t}}$ has been fixed at $-4.0$ TeV, leading to $m_{\tilde{Q}_3} = 1.95$ TeV. Here the value of the $\delta^{23}_{\tilde{u}LL}$ remains constrained by $B\rightarrow X_s \gamma$ and $B_s\rightarrow\mu^+\mu^-$. To some extent, these processes restrict the size of the product of $\delta^{12}_{\tilde{u}LL}\delta^{13}_{\tilde{u}LL}$. However, the only significant dependence is observed in the $\delta^{12}_{\tilde{u}LL}$ - $\delta^{13}_{\tilde{u}LL}$ case, where the effect of $\pm 5\%$ of the standard model branching ratio can be reached.

\section{Conclusions}

To conclude, we would like to point out that possible future observation of any non-standard flavor violating effect in $K^+\rightarrow\pi^+\nu\bar{\nu}$, up to $\pm 10\%$ in the decay branching ratio, can be interpreted in terms of the non-trivial flavor structure in the left-left squark sector. For this purpose, the large soft trilinear coupling $A_{\tilde{t}}$ as well as heavy gluinos have to be assumed. This scenario extends the list of the possible origins of such deviation from the standard model prediction, usually containing $\delta_{\tilde{u}LR}$ or $\delta_{\tilde{d}RR}$ only.

\section*{Acknowledgments}

We would like to thank Comenius University in Bratislava (grants UK/525/2011 and UK/64/2013) and Nad\'acia SPP (NSPP-Hlavi\v cka 2012/2013) for financial support.

\appendix
\section{Chargino \texorpdfstring{$\delta^{i3}_{\tilde{u}LL}$}{NMFV}-loops in mass insertion approximation}\label{A}

Full expressions corresponding to all the particular loop diagrams in Fig.~\ref{fig:MIAloops}, as they contribute to the $X_L$ coefficient, are presented here.
\begin{eqnarray}
X^{(a)}_L = \frac{V_{cd}}{V_{td}} \Delta^{i3*}_{\tilde{u}LL} A_{\tilde{t}}m^2_t M_2 \Big\{
-\tfrac{1}{2}\big(-\tfrac{1}{2}+\tfrac{1}{3}s^2_W\big)\big(E_0\big(.\big)+E_1\big(.\big) +E_2\big(.\big)\big)
\\+\big(1-s^2_W\big)F_{00}\big(.,M_2\big)-\big(\tfrac{1}{2}-s^2_W\big)\big(F_{00}\big(.,\mu\big) -\tfrac{1}{2}\mu^2 F_0\big(.,\mu\big)\big)\nonumber\\
-\tfrac{2}{3}s^2_W F_{00}\big(.,M_{\tilde{t}^c}\big)
+\big(\tfrac{1}{2}-\tfrac{2}{3}s^2_W\big)\big(F_{00}\big(.,M_{\tilde{u}_i}\big)+F_{00}\big(.,M_{\tilde{t}}\big)\big)
\Big\}\nonumber
\end{eqnarray}
\begin{eqnarray}
X^{(b)}_L = \frac{V^*_{cs}}{V^*_{ts}} \Delta^{23}_{\tilde{u}LL} A_{\tilde{t}}m^2_t M_2 \Big\{
-\tfrac{1}{2}\big(-\tfrac{1}{2}+\tfrac{1}{3}s^2_W\big)\big(E_0\big(.\big)+E_1\big(.\big)+E_2\big(.\big)\big)
\\+\big(1-s^2_W\big)F_{00}\big(.,M_2\big)-\big(\tfrac{1}{2}-s^2_W\big)\big(F_{00}\big(.,\mu\big) -\tfrac{1}{2}\mu^2 F_0\big(.,\mu\big)\big) \nonumber\\
-\tfrac{2}{3}s^2_W F_{00}\big(.,M_{\tilde{t}^c}\big)+\big(\tfrac{1}{2}-\tfrac{2}{3}s^2_W\big)\big(F_{00}\big(.,M_{\tilde{c}}\big)+F_{00}\big(.,M_{\tilde{t}}\big)\big)\Big\}\nonumber
\end{eqnarray}
\begin{eqnarray}
X^{(c)}_L = \frac{V^*_{is}V_{jd}}{V^*_{ts}V_{td}}\Delta^{i3}_{\tilde{u}LL} \Delta^{j3*}_{\tilde{u}LL}2 M^2_W M^2_2 \sin^2\beta \Big\{-\tfrac{1}{2}\big(-\tfrac{1}{2}+\tfrac{1}{3}s^2_W\big)\big(F_0\big(.\big)\\
+F_1\big(.\big)+F_2\big(.\big)+F_3\big(.\big)\big)+2\big(1-s^2_W\big)G_{00}\big(.,M_2\big)\nonumber\\
+\big(\tfrac{1}{2}- \tfrac{2}{3} s^2_W\big)\big(G_{00}\big(.,M_{\tilde{u}_i}\big) +G_{00}\big(.,M_{\tilde{t}}\big)+G_{00}\big(.,M_{\tilde{u}_j}\big)\big)\nonumber\\ 
-\big(\tfrac{1}{2}-s^2_W\big)\big(G_{00}\big(.,\mu\big)-\tfrac{1}{2}\mu^2 G_0\big(.,\mu\big) \big)\Big\}\nonumber
\end{eqnarray}
\begin{eqnarray}
X^{(d)}_L = -\frac{V^*_{is}V_{jd}}{V^*_{ts}V_{td}} \Delta^{i3}_{\tilde{u}LL} \Delta^{j3*}_{\tilde{u}LL}
A^2_{\tilde{t}}m^2_t\Big\{-\tfrac{1}{2}\big(-\tfrac{1}{2}+\tfrac{1}{3}s^2_W\big)
\big(F_0\big(.\big)+F_1\big(.\big)\big)\\
+\big(\tfrac{1}{2}-\tfrac{2}{3}s^2_W\big)\big(G_{00}\big(.,M_{\tilde{u}_i}\big)+2G_{00}\big(.,M_{\tilde{t}}\big)+G_{00}\big(., M_{\tilde{u}_j}\big)\big)\nonumber\\
-\tfrac{2}{3}s^2_W G_{00}\big(.,M_{\tilde{t}^c}\big)-\big(1-s^2_W\big)\big(G_{00}\big(.,M_2\big)-\tfrac{1}{2}M^2_2 G_0\big(.,M_2\big)\big)\Big\}\nonumber
\end{eqnarray}

In our notation $M^2_{\tilde{u}_i} = (M^2_{\tilde{u},LL})^{ii}$ and $M^2_{\tilde{u}^c_i} = (M^2_{\tilde{u},RR})^{ii}$. In each of this expressions, the ``$.$'' as an argument represents the set of masses of all particles participating in loop, as for example $. = \{M_{\tilde{u}_i},M_2,\mu,M_{\tilde{t}},M_{\tilde{t}^c}\}$ in $X^{(a)}_L$. This set is extended by the mass of the particle, to which internal leg the $Z$ propagator is attached. This can be seen also from the isospin and charge factors in brackets in front of the corresponding loop function. The definition of $E_0, \ldots$ can be found in Ref. \citen{Hahn:1998yk}.

\bibliographystyle{ws-ijmpa}      
\bibliography{kaon_decay}  

\end{document}